# Simultaneously Propagating Voltage and Pressure Pulses in Lipid Monolayers of pork brain and synthetic lipids

J. Griesbauer[1,2], S. Bössinger[1,2], A. Wixforth[1,3], M.F. Schneider[2]

[1] *University of Augsburg, Experimental Physics I, D-86159 Augsburg, Germany*

[2] *Boston University, Dept. of Mechanical Engineering, Boston-Massachusetts, USA*

[3] *Augsburg Center for Innovative Technologies (ACIT) and Center for NanoScience (CeNS)*

**Abstract**

Hydrated interfaces are ubiquitous in biology and appear on all length scales from ions, individual molecules to membranes and cellular networks. In vivo, they comprise a high degree of self-organization and complex entanglement, which limits their experimental accessibility by smearing out the individual phenomenology. The Langmuir technique, however, allows the examination of defined interfaces, whose controllable thermodynamic state enables one to explore the proper state diagrams.

Here we demonstrate that voltage and pressure pulses simultaneously propagate along monolayers comprised of either native pork brain or synthetic lipids. The excitation of pulses is conducted by the application of small droplets of acetic acid and monitored subsequently employing time-resolved Wilhelmy plate and Kelvin probe measurements. The isothermal state diagrams of the monolayers for both lateral pressure and surface potential are experimentally recorded, enabling us to predict dynamic voltage pulse amplitudes of 0,1 – 3mV based on the assumption of static mechano-electrical coupling.

We show that the underlying physics for such propagating pulses is the same for synthetic (DPPC) and natural extracted (Pork Brain) lipids and that the measured propagation velocities and pulse amplitudes depend on the compressibility of the interface. Given the ubiquitous presence of hydrated interfaces in biology, our experimental findings seem to support a fundamentally new mechanism for the propagation of signals and communication pathways in biology (signaling), which is neither based on protein-protein or receptor-ligand interaction nor on diffusion.

# Introduction

Lipid bilayers are ubiquitously entangled in biological processes, usually comprising the impact of changes in external conditions and the exposure to various substances. The static response of membranes to such conditions was extensively studied from a mechanical $(\pi)$ thermal $(T)$ and – although to a lesser extend - electrical $(\psi)$ perspective [1–13]. Non-equilibrium studies on the macroscopic dynamics of lipid membranes, on the other hand, are only sparsely found, to date [14–19]. The propagation of pulses along interfaces, however, would add a fundamentally new mechanism to the theory of inter and intra-cellular communication. Furthermore a proof of the existence of pulses propagating macroscopic distances would favorably support the idea of nerve pulse propagation in analogy to sound [20–24].

We have recently shown that acoustic pulses can indeed propagate over macroscopic distances in lipid monolayers [25], [26]. Such monolayers can serve as an excellent model system as they allow direct access to their physical parameters. For instance, a monolayer can be deposited on a water surface, while controlling the thermodynamic state by compression, pH-changes or by heating/cooling. At the same time, the response of the monolayer system can be monitored by measuring, for example, the lateral pressure $\pi$ with a Wilhelmy plate, giving access to the lateral compressibility $\kappa = \frac{-1}{A}\left(\frac{\partial A}{\partial \pi}\right)$, where $A$ is the occupied area of the monolayer. Moreover, a Kelvin probe allows simultaneous access to the so called surface potential $\psi$ of the lipid monolayer [2].

In this manuscript we study voltage pulses on lipid monolayers, which are accompanied by acoustic pressure pulses, with respect to the dynamic mechano-electrical coupling in lipid monolayers. Excited by small acetic acid droplets, the pulses are monitored in both lateral pressure $\pi$ and surface potential $\psi$ and find maximal pressure and potential amplitudes of ~ *0.3 mN/m* and *~3mV*, respectively. We show that *static* measurements of $\pi$ and $\psi$, on the other hand, allow the qualitative and quantitative calculation of dynamic pulse shapes and amplitudes in $\psi$ using pulse measurements in $\pi$. Our experiments demonstrate that both pulse amplitudes and velocities depend on the compressibility $\kappa$ and thereby on the thermodynamic state of the monolayer. These results are discussed in the framework of a simple linear hydrodynamic model, which correctly recovers our results.

**Materials and Methods**

Lipid monolayers of 1,2-dipalmitoyl-*sn*-glycero-3-phosphocholine (DPPC) or total lipid extract of pork brain (PBTE) were purchased from Avanti Polar Lipid (USA) and spread from Chloroform to the air/water-interface of a film balance trough [26], [27] (Fig. 1a). After *10 minutes of evaporation*, the lateral pressure and surface potential-area isotherm ($\pi$-A and $\psi$-A) are recorded by slow (~ *2,5Å$^2$/(min·molecule)*) compression of the film by means of a moveable barrier. The trough is equipped with two pressure sensors (Wilhelmy plates) and a Kelvin probe (vibrating capacitor method), which can be read out very rapidly (*10000 samples/second, 0.01 mN/m and 0.1 mV resolution*). Arriving pressure/potential pulses can thus be directly monitored by their mechanical and electrical responses, whereas the high sample rates allow for time resolved measurement and subsequent Fourier transform of the detected pulse shapes. As only the longitudinal pressure pulses within the monolayers are of interest, an additional barrier is introduced to exclude spurious effects of unwanted water waves (Fig. 1a). The actual $\pi-$ and $\psi-$pulse is excited in a separate compartment by sudden addition of a small amount of acetic acid (~ *3µl*) to the monolayer surface. Only pulses able to travel over macroscopic distances will cause the pressure sensors or the Kelvin probe to respond. To exclude artifacts by, e.g., unwanted water wave effects, we also performed reference measurements at which the acetic acid droplets were fused onto a pure water surface. Within the resolution of our experiments, we were not able to detect any recognizable pulse response during these reference measurements (see supplementary).

**Results**

Fig. 1b shows a typical result of a propagating $\pi-$ and $\psi-$pulse. The droplet was deposited onto a DPPC monolayer at *t ~ 1s* exciting a pulse. This pulse first arrives at pressure sensor 2, then at the Kelvin probe and eventually at pressure sensor 1 (Fig. 1a). The time delay between the two pressure sensor responses and their distance (~ *14,5cm*) can be used to directly extract the propagation velocity of the pulse travelling in the excitation compartment along the separation barrier into the detection compartment [25]. Fig. 1b demonstrates that along with the measured pressure *($\pi$)* pulses a nearly identically shaped pulse evolves in the surface potential $\psi$ of the monolayer. Indeed, theoretical considerations [27], [28] even predict that *all* thermodynamic observables change along with the surface pressure (or any other observable). This behavior is experimentally well reproduced for propagating pulses in $\pi$ and $\psi$ suggesting the idea of coupled observables also to be correct for dynamic processes.

To further evaluate the coupling of $\pi$ and $\psi$, their respective isotherms for a DPPC monolayer are displayed in Fig. 2a, together with the corresponding $\pi$-$\psi$-curve $\psi(\pi)$ (Fig. 2a inset). The latter was extracted by associating the values of $\pi(A)$ and $\psi(A)$ for matching areas $A$. For a pulse shape $\pi(t)$ in the lateral pressure, the function $\psi(\pi)$ can be used to predict the corresponding pulse shape in surface potential by $\psi(t) = \psi(\pi(t))$. For calculations, the function $\psi(\pi)$ is part-wise approximated linearizations with slopes as they are marked in the inset of Fig. 2a. Fig. 2b thus presents the predictions for the change in surface potential $\psi(t)$ based on the pulse $\pi(t)$ compared to the actual measurement of $\psi(t)$. Indeed, both prediction and measurement agree quantitatively and qualitatively without any fitting adjustment, rendering the usage of static measurement (inset of Fig. 2a) as valid.

Theoretically a purely adiabatic monolayer pulse, being decoupled from its viscous water sub phase, may be described by the one dimensional, classical wave equation [26], [29].

$$\frac{\partial^2 v_m}{\partial t^2} - c_0^2 \frac{\partial^2 v_m}{\partial x^2} = 0, \ where \ c_0 = \frac{\sqrt{1}}{\rho_0 \kappa_S} \quad (1)$$

Here, $v_m$ is the velocity field of the monolayer, $\rho_0$ its lateral density and $\kappa_S$ its adiabatic compressibility, respectively. The latter will be approximated by the isothermal lateral compressibility $\kappa_T$, which can be directly extracted from measured $\pi$-$A$-isotherms.
Theoretical evaluations of $c_0$ using $\kappa_T$ for a DPPC monolayer yield $c_0 \sim 50\text{-}200 \ m/s$, which is in contradiction to the pulse propagation velocities of $c \sim 1 \ m/s$, experimentally determined here. One way to account for this discrepancy is to include a coupling of the monolayer lipids to their water sub phase. In the simplest assumption of direct coupling of the monolayer to the aqueous sub phase the application of the Stokes-Equations to this system leads to the following extended wave equation (for a detailed derivation see [25]):

$$\frac{\partial^2 v_m}{\partial t^2} + \frac{1}{\rho_0} e^{i\frac{\pi}{4}} \sqrt{\eta_w \rho_w \omega} \frac{\partial v_m}{\partial t} - c_0^2 \frac{\partial^2 v_m}{\partial x^2} = 0 \quad (2)$$

Here, $\eta_W$ and $\rho_W$ represent the water viscosity and density, respectively, while $\omega$ denotes the pulse's mean wave frequency. We have chosen this monolayer motivated perspective over a capillary driven approach, due to the clear dependence of the propagation velocity on the elastic properties of the film. Although our approach matches the data excellently, it should be noted that in an capillary wave base theory the system can be treated as a free interface with a adsorbed compressible film, too [14, 15]. We currently discussing such an approach with collaborators from theory and will report the results properly elsewhere.

However, since in the observed pulse shapes, as shown in Fig. 1b, imply low frequencies of approximately ~ *1Hz* the resulting propagation velocity *c* in equation (2) can be well approximated b:

$$c = \frac{\omega}{R(k)} = cos^{-1}\left(\frac{\pi}{8}\right)\sqrt{\frac{1}{\kappa_S}\frac{\sqrt{\omega}}{\eta_w \rho_w}} \quad (3)$$

In Fig. 3a, pulse propagation velocities, which were extracted from the run time of the pressure signal between the two sensors, are shown as solid lines for a DPPC monolayer at *24°C*. Dotted lines represent the calculated results from Equ. (3) using the independently evaluated isothermal compressibility $\kappa_T = \frac{-1}{A}\left(\frac{\partial A}{\partial \pi}\right)_T$ and three different frequencies $\omega \sim$ *1 Hz / 9 Hz / 18 Hz*. Both velocities are plotted as a function of $\pi$ and reveal a very good agreement. Considering the general constraint $\kappa_S < \kappa_T$ [31], it follows that the directly measured propagation velocities (from $\kappa_S$) have to be faster than the velocities extracted from Equ. (3) using $\kappa_T$. Therefore, Fig. 3a implies pulse frequencies of $\omega \sim$ *1 Hz* for $\pi <$ *10 mN/m* and frequencies of at least $\omega \sim$ *9 Hz* for $\pi >$ *10 mN/m*. Employing a Fourier transform of the pulse shapes reveals a very similar frequency range.

Identical measurements on cell-extracted lipid (PBTE) monolayers confirm the correlation of propagation velocity and compressibility as implied by Equ. (3) (Fig. 4a). At the same time it is remarkable that PBTE monolayers also indicate frequencies of $\omega \sim$ *1 Hz* for $\pi <$ *10 mN/m* and frequencies of at least $\omega \sim$ *9 Hz* for $\pi >$ *10 mN/m*. This would characterize the pulse frequencies to be pressure dependent. Apart from this the correlation of surface potential and lateral pressure isotherms (Fig. 5a) indicate the height of surface potential pulses on PBTE monolayers. As can be seen at the scale of Fig. 5b the height of PBTE potential pulses is one order of magnitude smaller than those of DPPC monolayers.

Nevertheless the measurements on PBTE monolayers demonstrate the biological relevance, showing, that the theory of pulse propagation in synthetic monolayers is also applicable to biological systems.

For the sake of completeness, the excitability (pulse heights) of DPPC monolayers (Fig. 3b)

is shown in Fig. 3b. Similar to the propagation velocities, the peak heights show a dependence following the compressibility of the monolayer and therefore its thermodynamic state. Indeed, this behavior even applies to PBTE (see Fig 4b) and therefore utterly underlines the general coupling of all observables via the thermodynamic state.

**Conclusion**

Our results demonstrate that voltage-pulses in the low *mV* range, which are inevitably coupled to acoustic pressure-pulses, can propagate along lipid monolayers. Thermodynamic couplings known from static–isothermal experiments are therefore also found under non-equilibrium conditions. Given the proper detection, we conclude from our results that a corresponding pH and temperature pulse must exist as well, although we expected the latter to be small. As the lipid bilayer follows – at least qualitatively - the same physics as the monolayer (under certain conditions), these results are in support of a thermodynamic foundation of the nervous impulse [20], [21], [24]. More importantly, they provide a basis to propose a new mechanism of inter- and intracellular communication in biology (e.g. signaling) in general [25]. *Wherever interfaces exist*, which are locally in contact to another system (e.g. a bath, a substance, etc.), *a finite probability for (spontaneous or controlled) pulse excitation is expected*. A protein, for instance, embedded in a membrane would "experience" the transient collective changes of the interface (e.g. compression, electric field, etc.) and react accordingly. Exciting candidates are enzymes, which are known to exhibit a very strong coupling between activity and interfacial state [31–33]. Whether the mechanical ($\pi$-A) or electrical diagrams of state determine the coupling between propagating induced state and enzyme activity, however, depends on the mechanical and electrical properties of the single molecule. In any case, we believe that our studies predict the propagation of pulses in biological interfaces and suggest a novel alternative way of communication and signaling between biological entities on scales ranging from individual enzymes to entire membrane complexes.

. .

**Figure 1 a)** Experimental setup of the film balance used for Monolayer-Pulse-Analysis. The balance trough is equipped with two Wilhelmy type pressure sensors and a Kelvin probe, such that both lateral pressure and surface potential can be recorded time-resolved. The additional barrier separates the excitation from the detection site to ensure the suppression of spurious water waves. **b)** Time course of the simultaneous readout of all three sensors after pulse excitation on a DPPC monolayer (*24°C*). The signals arrive at the sensors according to Fig. 1 a), such that the pulse travels from pressure sensor 2, to the Kelvin probe, to pressure sensor 1.

**Figure 2 a)** Isothermal, quasistatic recording of both lateral pressure $\pi$ and surface potential $\psi$ of a DPPC monolayer (*24°C*). Both isotherms exhibit a flat plateau, indicating the phase transition from the liquid-expanded to the liquid-condensed phase, whereas the initial formation (before the liquid-expanded phase) is indicated by the first rises of the surface potential. In the inset, we correlate surface potential and lateral pressure of the same lipid areas resulting in a Potential-to-Pressure plot. **b)** Time course of surface potential (green) and lateral pressure (red) recorded for a traveling pulse at the detection site. Using the correlation between pressure and potential as indicated by the inset of Fig. 2 a), the pressure course $\pi(t)$ is used to calculate a prediction for the potential $\psi(t)$ (blue).

**Figure 3 a)** Propagation velocities of the pulses excited by acetic acid. On the one hand, the velocities are extracted by the runtime difference between the two pressure sensors (distance ~*14,5cm*) for different lateral pressures of DPPC monolayers (*24°C*). On the other hand, the isothermal compressibility $\kappa_T$, as shown in the inset of Fig 3 b), is used in the model of Equ. (3) and plotted for three different pulse frequencies of ~*1Hz, ~9Hz, ~18Hz*. Indeed, the coincidence of model and measurement indicates frequencies of ~*1Hz* for low lateral pressures (< *10 mN/m*) and frequencies of ~*9Hz* for high lateral pressures (> *10 mN/m*). **b)** Measured pulse amplitudes for different lateral pressures of the DPPC-Monolayer (*24°C*). Similar to the propagation velocities the amplitudes follow the phase state of the monolayer indicated by $\kappa_T$.

**Figure 4 a)** Propagation velocities of the pulses excited by acetic acid on PBTE-Monolayers (*24°C*). Both range of velocities and relation to compressibility demonstrate the same underlying physics as in figure 3a) showing that this is a general behavior independent of the respective lipid compositions. **b)** Measured pulse amplitudes for different lateral pressures of the PBTE-Monolayer (*24°C*). As expected, the velocity exhibits neither minima nor maxima, since the corresponding compressibility $\kappa_T$ of PBTE monolayers has no distinct maxima or minima.

**Figure 5 a)** Isothermal, quasistatic recording of both lateral pressure $\pi$ and surface potential $\psi$ of a PBTE monolayer (*24°C*). In contrast to the DPPC isotherms of Figure 2 a) no plateau region can be observed. In the inset, we again correlate surface potential and lateral pressure of the same lipid areas resulting in a Potential-to-Pressure plot. **b)** Time course of lateral pressure (red) recorded for a traveling pulse at the detection site. Using the correlation between pressure and potential as indicated by the inset of Fig. 5 a), the pressure course $\pi(t)$ is used to calculate a prediction for the potential $\psi(t)$ (blue).

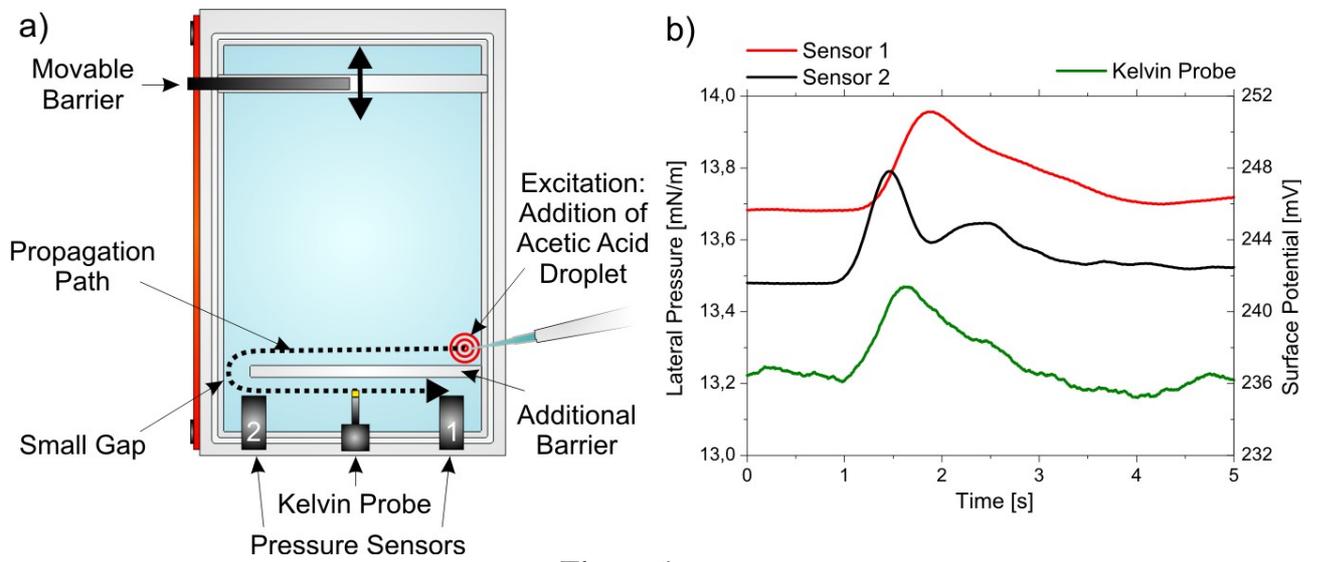

**Figure 1**

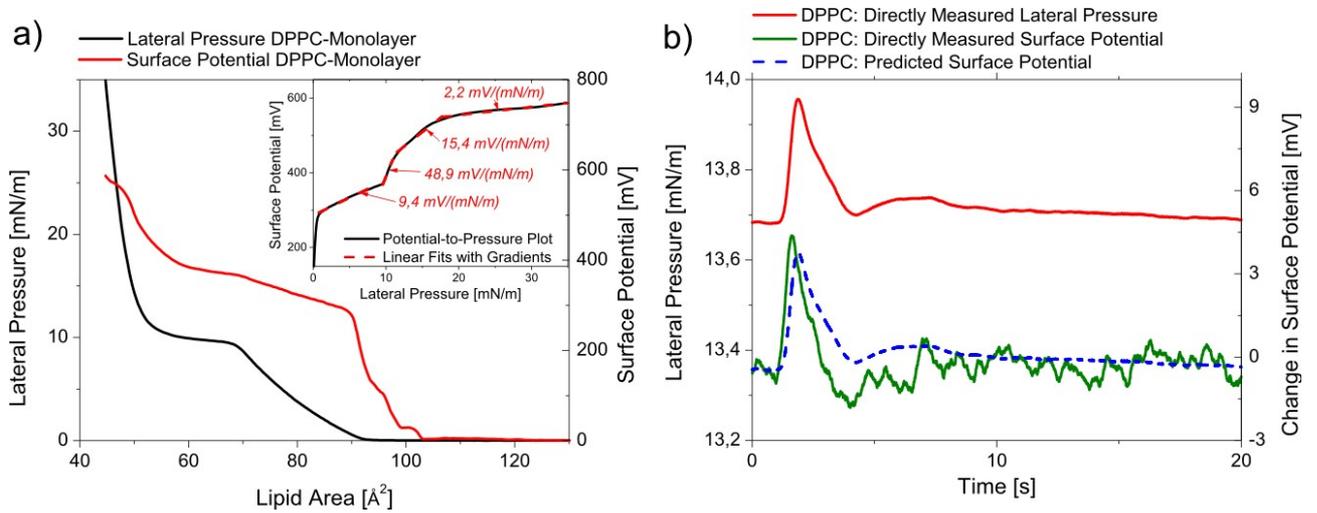

**Figure 2**

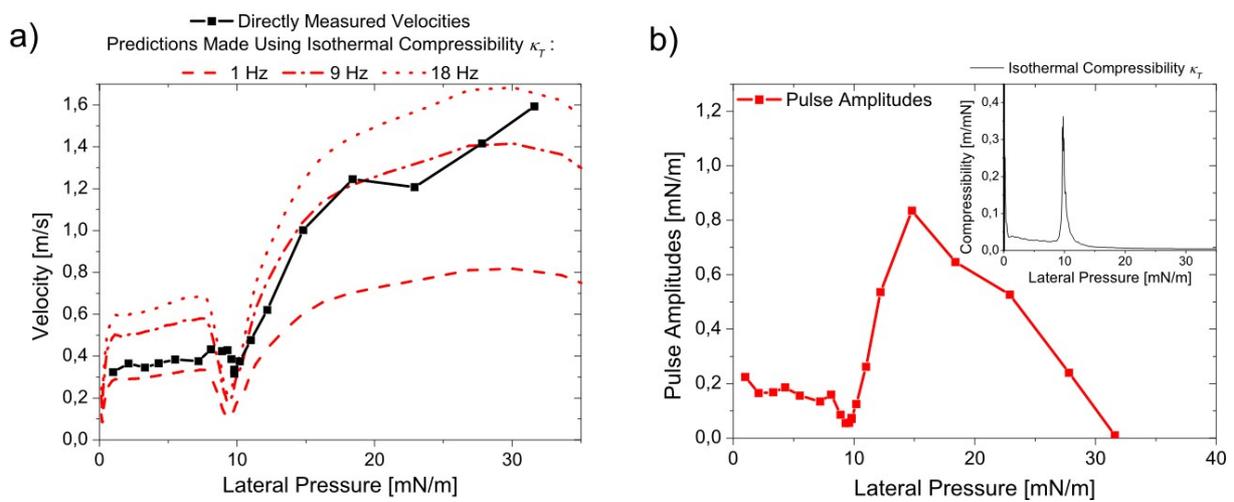

**Figure 3**

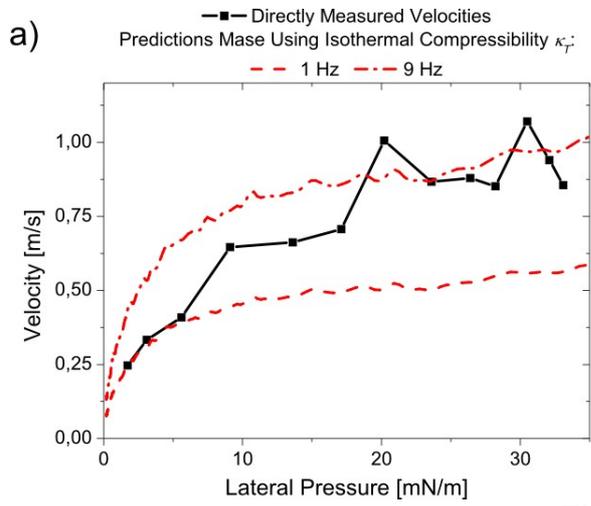 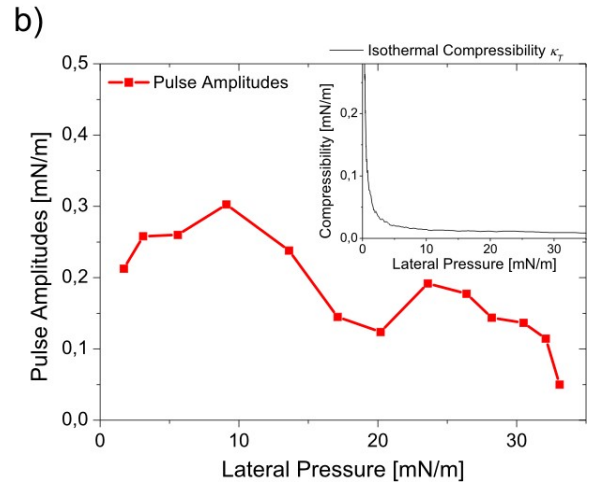

**Figure 4**

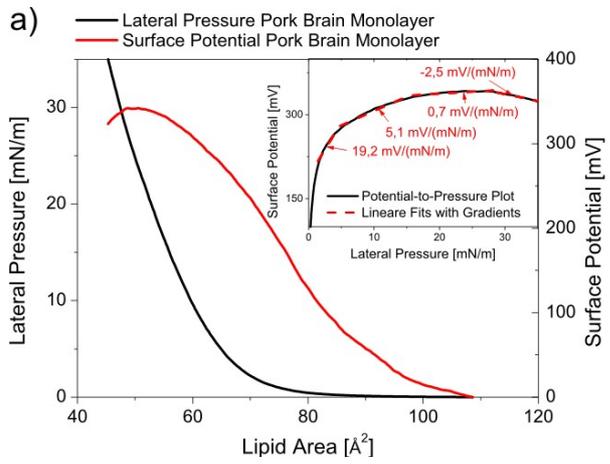 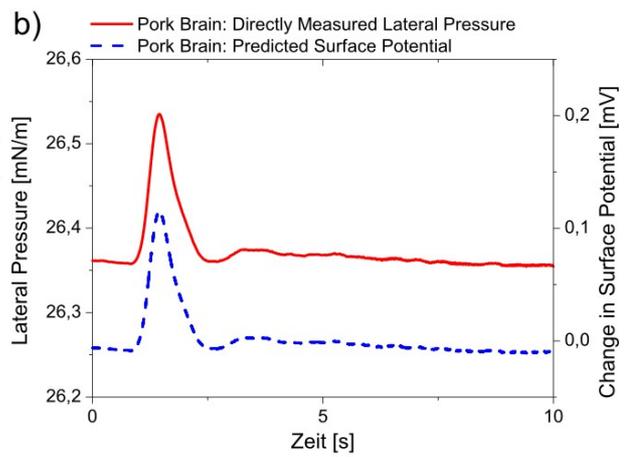

**Figure 5**